\documentclass[twocolumn,showpacs,preprintnumbers,amsmath,amssymb,superscriptaddress]{revtex4}

\usepackage{graphicx}
\usepackage{dcolumn}
\usepackage{bm}
\usepackage{multirow}%

\usepackage{color}

\newcommand{\half}{\frac{1}{2}}
\newcommand{\Ecnl}{E^{\rm c}_{\rm nl}}
\newcommand{\EvdWO}{E^{\rm vdW0}}
\newcommand{\Eads}{E_{\rm ads}}

\begin{document}

\title{Influence of van der Waals Forces on the Adsorption Structure of
Benzene on Silicon} 
\author{Karen Johnston}
\affiliation{Laboratory of Physics, Helsinki University of
  Technology, P.O. Box 1100, 02015, Finland}
\author{Jesper Kleis}
\affiliation{Center for Atomic-scale Materials Design, Technical University of Denmark, DK-2800 Kgs. Lyngby, Denmark}
\author{Bengt I. Lundqvist}
\affiliation{Center for Atomic-scale Materials Design, Technical University of Denmark, DK-2800 Kgs. Lyngby, Denmark}
\affiliation{Department of Applied Physics, Chalmers University
  of Technology, Gothenburg, SE-41296, Sweden}
\author{Risto M. Nieminen}
\affiliation{Laboratory of Physics, Helsinki University of
  Technology, P.O. Box 1100, 02015, Finland}

\begin{abstract}
Two different adsorption configurations of benzene on the
Si(001)-(2$\times$1) surface, the tight-bridge and butterfly
structures, were studied using density functional theory.
Several exchange and correlation functionals were used, including
the recently developed vdW-DF functional, which accounts for the
effect of van der Waals forces.  In contrast to the PBE, revPBE
and other GGA functionals, the vdW-DF functional finds that, for
most coverages, the adsorption energy of the butterfly structure
is greater than that of the tight-bridge structure.   
\end{abstract}
\maketitle

In the quest for reduced-sized transistor chips the combination
of organic molecules with silicon-based technology is of
increasing importance.   
The ability to manipulate organic molecules on surfaces is
developing rapidly and an understanding of the structural and
transport properties of adsorbed molecules is essential
\cite{Hersam2000a,Bent2002a,Duwez2006a,Zanella2006a,Quek2007a}.   

Silicon is not only important for technology but also
demonstrates the versatility of the covalent bond. Covalency
makes bulk silicon, diamond, and graphene layers very strong.  At the
same time, it can produce a multitude of competing atomic
structures when spatial restrictions are released. For instance,
the Si(111) and Si(001) surfaces show quite different properties,
including radically different types of reconstructions. While the
covalent bond is typically very strong, the energy differences
between such reconstructions can be small \cite{Stich1992a}.
Their relative
stabilities and the influence from adsorbates are interesting
issues to understand. In particular, there may be situations where
the weak van der Waals (vdW) force can significantly influence
the structures. 
The benzene molecule demonstrates several types of internal bonds,
typically interacts with other molecules via vdW forces, and is an
important model unit for several classes of large molecules like
DNA \cite{Cooper2007b}.  The adsorption of 
benzene on Si is obviously very interesting, both because of the
versatility in bond types and the wide ramifications, including
technological ones, such as molecular electronics.  

\begin{figure}[ht!]
\begin{center}
\includegraphics[height=2.4cm]{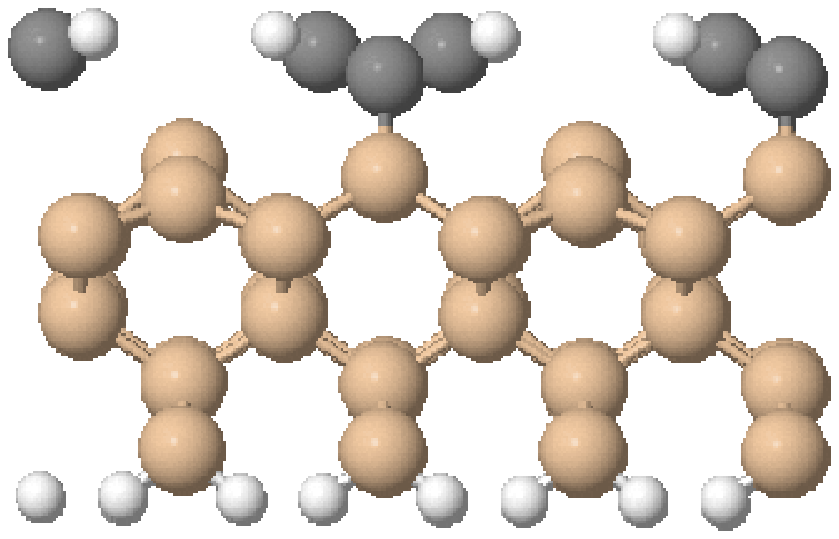}
\includegraphics[height=2.4cm]{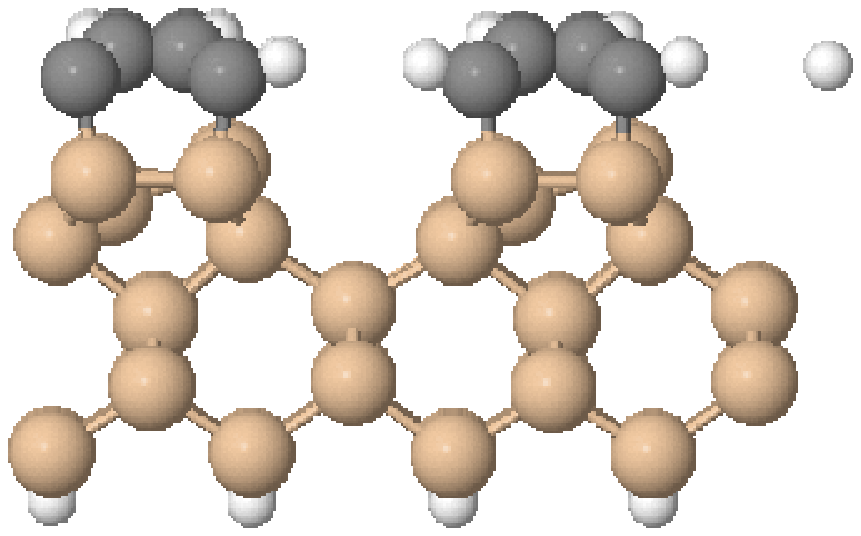}
\end{center}
\begin{center}
\includegraphics[height=2.4cm]{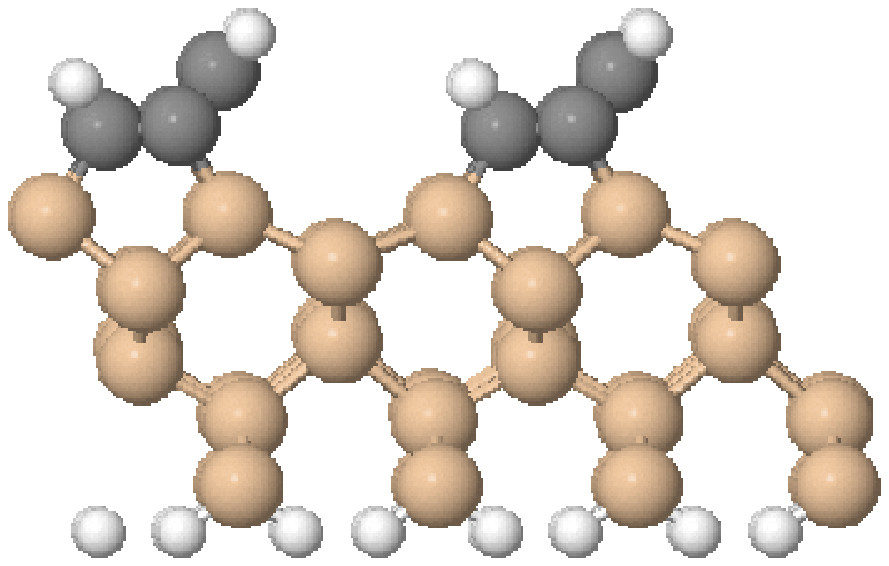}
\includegraphics[height=2.4cm]{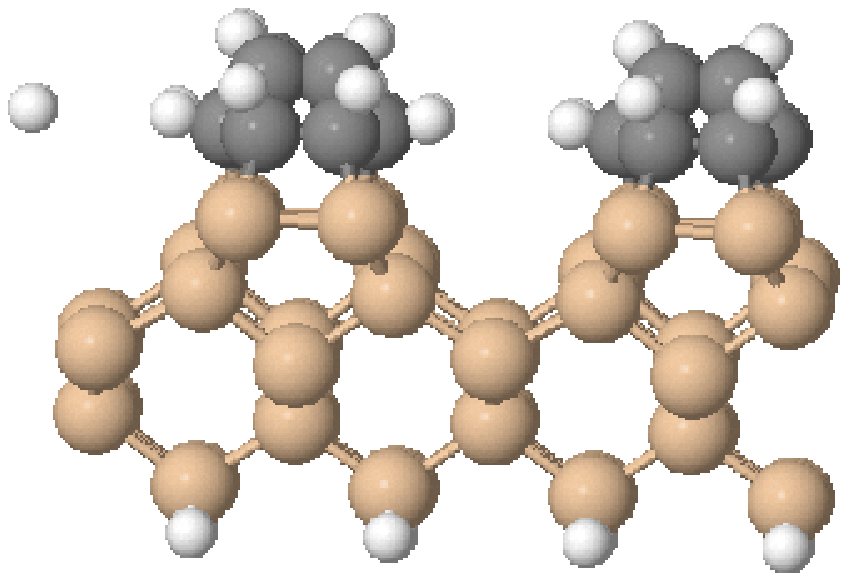}
\caption{\label{fig:BF_TB} (Color online) Butterfly (BF) (top),
  and tight-bridge (TB) (bottom) structures of C$_6$H$_6$ on
Si(001)-(2$\times$1).     
LHS: viewed along [$\bar{1}10$].  RHS: viewed along [110] dimer
  rows.}
\end{center}
\end{figure}

Adsorption of benzene on the Si(001)-(2$\times$1) surface has been
studied extensively, however, without unanimous results.  There
is agreement on the two most stable structures, shown in
Fig.~\ref{fig:BF_TB}, but to date
there is no agreement which is the stable one.  
The butterfly (BF) structure is adsorbed on top of a single dimer and has two
symmetry planes: along and perpendicular to the dimer row.
The tight-bridge (TB) structure is adsorbed across two dimers and has one
symmetry plane along the dimer row.  

To differentiate between the adsorption structures and their
bonding and symmetry properties, a variety of experimental tools
have been employed.  
Thermal desorption and angle-resolved photoelectron
spectroscopy were used to investigate the electronic structure and
symmetry of benzene on Si \cite{Gokhale1998a} and a single dimer
structure was observed, supporting the BF configuration.  
Near-edge X-ray-absorption fine-structure found the benzene to be
symmetric with respect to the dimer axis, ruling out the TB
structure \cite{Witkowski2003a}.  These findings were supported
by optical spectroscopy data \cite{Witkowski2005a}, which found
that benzene adsorbs on top of a single dimer rather than on the
bridge site between two dimers.  
However, according to scanning tunnelling microscopy (STM)
studies \cite{Lopinski1998a,Lopinski1998b}, the benzene
molecule adsorbs initially in the BF structure but this is
observed to be metastable with respect to a bridging
configuration.  With the STM tip one benzene structure can be
converted to another, with the conversion-energy barrier
estimated to be 0.95~eV.  
A high-resolution study \cite{Kim2005a} suggested that the
adsorption geometry depends on coverage \footnote{In this paper
one monolayer (ML) corresponds to one benzene molecule per Si
dimer.} and showed a bridging structure is favored at low
coverages while at high coverages a single dimer structure (BF)
is more stable.  This would explain the discrepancy between the 
experimental results of
Refs.~\onlinecite{Gokhale1998a,Witkowski2003a,Witkowski2005a} 
that supported the BF structure, as they were performed at
saturation coverage, while the STM experiments were carried out
at a low coverage of 0.044~ML.  

Adsorption energies, $\Eads$, have been calculated using standard density
functional theory (DFT) and all of these studies
\cite{Hofer2001a,Lee2005a,Mamatkulov2006a} support the TB structure.  
An MP2 (second order perturbation/quantum mechanics molecular
dynamics) cluster calculation \cite{Jung2005a} is the only
calculation which supports the BF structure ($\Eads$=1.04~eV).
Unlike standard DFT calculations, MP2 methods include vdW forces
but because they are computationally heavy only small systems can
be treated with this method.  Si (001) is an extended material
and it is not clear whether small clusters can give accurate
adsorption energies.
Furthermore, the cluster geometry corresponds to a low coverage
situation so cannot be compared with saturation coverage results. 

The vdW-DF functional \cite{Dion2004a} was developed to account for
the effect of vdW forces in DFT.  It has been shown to give
accurate results for molecular systems, such as benzene dimers
\cite{Puzder2006a,Thonhauser2006a}, and recently it has been 
successfully applied to systems with covalent bonding
present \cite{Thonhauser2007a,Ziambaras2007a,Chakarova2006a}.  
It is currently believed that vdW forces are only important in
physisorbed systems but in this paper we demonstrate that this is
not the case.  By using vdW-DF in DFT calculations we show that
in most of the studied configurations vdW forces stabilise the BF
structure, which is surprising since the TB structure has more 
covalent bonds and would be expected to be more strongly bound to
the surface. 

Standard DFT calculations were performed with Dacapo \cite{Dacapo} using the
PBE form \cite{Perdew1996a,Perdew1997a,Perdew1998a} of the
generalised-gradient approximation (GGA) and ultrasoft (US) 
pseudopotentials.  The plane-wave energy cutoff was
400~eV and the Brillouin zone mesh used was equivalent to
4$\times$4$\times$1 Monkhorst-Pack $k$-point sampling for
0.5~ML coverages. 
We used 9-atomic layer Si slabs, $\approx$15~{\AA} of vacuum and
a lattice constant of 5.47~{\AA}.  To save computational time
adsorption on only one side of the slab was considered.  The
bottom layer of the slab was fixed in the bulk Si positions and
passivated with two H atoms per Si atom.  The slabs were
electrostatically decoupled along the $c$-direction.  All
relaxations were considered complete when the forces were less
than 10~meV{\AA}$^{-1}$.   

Chemisorption energies are known to be sensitive to the choice of
the exchange and correlation (xc) functional.  It has been shown
that for chemisorbed atoms and molecules on transition metal
surfaces the revPBE functional \cite{Zhang1998a} generally gives
better results than the PBE functional, which tends to overbind
the molecules \cite{Hammer1999a}.  
A comparison between the present $\Eads$ and those of
previous studies is shown in Table~\ref{tab:DFT}. 
The PBE and PW91 \cite{Perdew1991a,Perdew1996b} xc functionals
give similar $\Eads$, as expected, but are around two
times higher than the revPBE $\Eads$.  
\begin{table}[ht!]
\begin{center}\begin{tabular}{rr|lll}\hline \hline
\multicolumn{2}{r|}{$\Eads$(eV/molecule)} & & &  \\
TB   & BF   & Pseudopotential & GGA      & Reference \\ \hline \hline
0.66 & 0.47 & US    & revPBE  & Present \\
1.16 & 0.89 & US    & PBE     & Present \\
1.24 & 0.99 & US    & PW91    & Present \\ \hline
1.25 & 1.00 & PAW   & PW91    & \cite{Johnston2007a} \\
1.05 & 0.82 & US/NC & PBE     & \cite{Lee2005a} \\
\hline
\end{tabular}
\caption{\label{tab:DFT} $\Eads$ of benzene in the TB
  and BF geometries for a coverage of 0.5~ML.  The
  pseudopotentials (psp) and xc functionals used are also shown.  }
\end{center}
\end{table}

The additional binding energy due to vdW forces is calculated
using the post-GGA total energy vdW  density functional (vdW-DF)
described in Ref.~\onlinecite{Dion2004a}.  To minimize any
artificial exchange binding which can be mistaken from
vdW-binding, the vdW-DF employs the revPBE form for the exchange
description \cite{Wu2001a,Langreth2005a}.   
The vdW-DF replaces the GGA correlation and divides the correlation into a
shorter ranged and a longer ranged part. The first part is approximated by
the LDA, while the latter ($\Ecnl$) includes the important dispersive
interactions.  $\Ecnl$ is nonlocal by construction and, consistent with the
approximation for the shorter ranged correlation, it is constructed to
vanish for a homogeneous system.
In the vdW-DF the total energy reads:
\begin{eqnarray}
\label{eqn:vdW-DF}
E^{\rm vdW-DF}
 &=& E^{\rm revPBE} - E^{\rm revPBE}_{\rm c} + E^{\rm LDA}_{\rm c} +  \Ecnl\\ \nonumber
 &=& \EvdWO + \Ecnl,
\end{eqnarray}
\noindent
with all terms obtained on the basis of self-consistent semi-local PBE DFT calculations. 
The nonlocal
correlation can be written in the simple form \cite{Dion2004a}:
\begin{equation}
 \Ecnl
 = \half \int {\rm d}{\bf r}\int {\rm d}{\bf
 r^\prime}
n({\bf r})\phi({\bf r},{\bf r^\prime})n({\bf r^\prime}),\label{eq:ecnl}
\end{equation}
with a density-density interaction kernel, $\phi({\bf r},{\bf
r^\prime})$ derived from the many-body response of the weakly
inhomogeneous electron gas.  
The kernel $\phi({\bf r},{\bf r^\prime})$ can be tabulated in
terms of two dimensionless parameters, a scaled separation
$D=\left|{\bf r}-{\bf r^\prime}\right|(q_0+q_0^\prime)/2$ and an 
asymmetry parameter $\delta=(q_0-q_0^\prime)/(q_0+q_0^\prime)$,
where $q_0({\bf r})$ is a local parameter that depends on the
electron density and its gradient at ${\bf r}$.
More information on the explicit form and deriviation of the
kernel can be found in Ref.~\onlinecite{Dion2004a}. 

$\Ecnl$ is to be evaluated for a periodic system and we use the
scheme developed in \cite{Kleis2007a,Ziambaras2007a} to evaluate
the integral. In short, we let  $\mathbf r$ in
Eq.~(\ref{eq:ecnl}) run through all electron density points
within the DFT unit cell.  The primed spatial coordinate on the
other hand, includes the electron density from the surrounding
lateral repeated images until convergence is obtained. We have found
the integral to be converged to include in total 5 (3) unit cells in each
direction for the largest (lowest) coverages. 

The post-GGA version of the vdW-DF functional implemented in this
article does not allow for any additional geometric and
electronic relaxation beyond the PBE determined adsorption
structure. 
Thonhauser {\it et al} \cite{Thonhauser2007a} implemented the
vdW-DF functional self-consistently and for noble gas and
cytosine dimers the difference between the self-consistent and
non-selfconsistent energies was negligible. This study also 
looked at the effect of the vdW-DF functional
on bulk Si.  In this case, as expected, the contribution of the
$\Ecnl$ energy was negligible and the lattice constant and total
energy were similar to those obtained using PBE.  
This indicates that our results, which use the non-selfconsistent 
implementation, are reliable. 

Due to numerical convergence issues of the nonlocal integral
\cite{Chakarova2006a,Ziambaras2007a}, special care is taken
when correcting for the nonlocal energy in the adsorption system,
leading to additional steps in the evaluation procedure of
vdW-DF. First, $\Eads$ between the benzene layer and 
the surface (BLS) is evaluated, and second, the intralayer
molecular-molecular (IMM) energy associated by isolating the
benzene from the benzene layer is calculated.  To ensure maximal
error cancelation in the evaluation of the above intermediates,
the isolated layer, the silicon surface and the benzene
molecule are fixed in the PBE adsorbant atomic configurations and
the spatial separation of the density fast-fourier transform grid
is kept constant in all calculations. Finally, the
contribution of unfolding (UF) the isolated molecule and the
surface to their PBE relaxed structures is calculated. We choose
in accord with Ref.~\cite{Chakarova2006a} to describe this final
energetic contribution within the PBE approximation. This is
justified, as all structural information has been obtained
self-consistently within the PBE functional, and furthermore, PBE
is, in contrary to vdW-DF, designed with the energetics of the
internal atomic binding in mind. 

The detailed data for the 0.5 ML case are given in Table~\ref{tab:vdW-DF}, with
the vdW-DF total energy separated into the nonlocal correlation
$\Ecnl$ including dispersion forces and the remaining $\EvdWO$ part.
\begin{table}[ht!]
\begin{center}
\caption{\label{tab:vdW-DF} vdW-DF $\Eads$ and
  its contributions, are shown for a coverage of 0.5~ML. The
  standard DFT PW91, PBE and revPBE results are shown for comparison.
  All energies are in eV. } 
\begin{tabular}{l|rrrr|rr} \hline \hline
             & PW91    & PBE     & revPBE  & vdW-DF &$\EvdWO$& $\Ecnl$ \\ \hline \hline
BLS (BF)     &    3.51 &    3.45 &    3.03 &    3.32 &    1.96& 1.36 \\
BLS (TB)     &    8.51 &    8.45 &    7.98 &    8.02 &    6.56& 1.46 \\ \hline
IMM (BF)     &    0.00 & $-$0.03 & $-$0.05 &    0.02 & $-$0.04& 0.06 \\
IMM (TB)     & $-$0.03 & $-$0.06 & $-$0.07 & $-$0.02 & $-$0.07& 0.05 \\ \hline
UF (BF)      & $-$2.52 & $-$2.52 & $-$2.51 &($-$2.52)&     ---&  --- \\
UF (TB)      & $-$7.23 & $-$7.23 & $-$7.24 &($-$7.23)&     ---&  --- \\ \hline
$\Eads$ (BF) &    0.99 &    0.89 &    0.47 &    0.82 &     ---&  --- \\
$\Eads$ (TB) &    1.24 &    1.16 &    0.66 &    0.77 &     ---&  --- \\ \hline \hline
\end{tabular}
\end{center}
\end{table}
We will briefly discuss both of the vdW-DF contributions that
contain nonlocal components (BLS and IMM) for the 0.5~ML case, and
compare these to the corresponding PBE and revPBE results.
 
{\it BLS interactions:}  $\Ecnl$ for the TB case
is found to be 0.1~eV more attractive than in the BF case. This
stems from the fact that the TB structure is more closely bound
to the Si surface, and accordingly we find $\approx$0.1~eV
difference for all coverages. However, $\Ecnl$ must be added
to $\EvdWO$ to find the vdW-DF energy. The $\EvdWO$ of the TB
configuration is at first glance considerably larger in the BF
case. However, comparing the combined vdW-DF energy with the
corresponding revPBE value, we see that the TB case has almost
the same energy, while the BF case is about 0.24~eV more
attractive.  Thus, compared to revPBE correlation energy the BLS
part of the vdW-DF correlation increases the BF $\Eads$
by $0.26$~eV more than in the TB case.  This is in accord 
with the expectation that the BF structure has a larger
vdW-like interaction than the more covalently bound TB structure.  

{\it IMM interactions:} We find $\Ecnl$ to be small and
attractive ($\approx0.05$~eV) while $\EvdWO$ is slightly
repulsive in both the TB and the BF configurations.  The IMM
$\EvdWO$ is found to be slightly more repulsive for the TB
configuration than for the BF, such that the net vdW-DF energy is
slightly repulsive for the TB case and weakly attractive for the
BF case, closely resembling the interactions obtained in the
semi-local revPBE and PBE approximations.  As the coverage is
reduced, the vdW-DF IMM interactions are even less pronounced and
can be described with the semi-local DFT functionals to within
0.02~eV.  

Overall, the change in $\Eads$ calculated with vdW-DF
is substantial, in particular for the BF structure.  The main
contribution to $\Eads$ comes from $\Ecnl$ between
the surface and the benzene layer.  The attraction between the
benzene molecules in the layer is minimal.  This behaviour is
similar to that of a cytosine dimer \cite{Thonhauser2007a} where
the repulsive $\EvdWO$ term is compensated by the attractive
$\Ecnl$ to give an overall binding energy of around 0.3~eV.  

We will now discuss the effect of coverage on the adsorption
energy.  The adsorption energies for various coverages 
are shown in Table~\ref{tab:coverage}.
\begin{table}[ht!]
\begin{center}
\caption{\label{tab:coverage} Variation of adsorption energy
  with coverage for the BF and TB structures.  The PW91, PBE and
  revPBE results are also shown for comparison.  All energies are in
  eV.} 
\begin{tabular}{lrrrr}\hline \hline
         & PW91 & PBE & revPBE & vdW-DF\\ \hline \hline
BF-0.5   & 0.99 & 0.89 & 0.47 & 0.82 \\
BF-0.25a & 1.02 & 0.93 & 0.51 & 0.82 \\
BF-0.25b & 1.02 & 0.93 & 0.61 & 0.84 \\
BF-0.125 & 1.04 & 0.96 & 0.54 & 0.84 \\ \hline
TB-0.5   & 1.24 & 1.16 & 0.66 & 0.77 \\
TB-0.25a & 1.23 & 1.16 & 0.67 & 0.74 \\
TB-0.25b & 1.33 & 1.25 & 0.76 & 0.86 \\
TB-0.125 & 1.31 & 1.24 & 0.75 & 0.82 \\ \hline
\end{tabular}
\end{center}
\end{table}
Both PBE and revPBE favor the TB configuration and, as
expected, PBE predicts the larger adsorption energy.  On the
contrary, vdW-DF, with its 
account of the dispersion interactions, predicts the BF
configuration to be slightly favored.  
For a coverage of 0.25~ML two supercell orientations are possible
and are denoted (a) and (b) with primitive lattice vectors
(220)($\bar 1$10)(006) and (110)($\bar 2$20)(006),
respectively.  In general only a minimal coverage dependence is
found as changes in $\Ecnl$ are almost cancelled by the
corresponding changes in $\EvdWO$. Correspondingly, the vdW-DF
adsorption energies in the BF case only show a small increase
with increasing coverage.  For the TB structure, the coverage
dependence is almost constant except for 0.25b coverage, which
has a pronounced preference.  This effect is also observed for the
semi-local functionals.  For all coverages the semi-local
functionals favor the TB configuration.  In contrast, vdW-DF
predicts that the BF structure is stable for all cases,
except 0.25b where the TB is lower in energy by only 0.02~eV. 

In the 0.125~ML case the nonlocal IMM interactions are less
than 0.01~eV so the monomers can be regarded as
isolated.  The difference between the BF and TB
configurations is thus 0.02~eV in the isolated case, which is
slightly less than the 0.08~eV difference found in a
corresponding MP2 calculation \cite{Jung2005a}. 
To compare with experimental data we used the Redhead
equation \cite{Redhead1962a} to estimate the adsorption
energies based on thermal desorption spectra in
Ref.~\cite{Gokhale1998a}.  For a peak temperature of 432~K, a
heating rate of 5~Ks$^{-1}$ and a pre-exponential frequency
factor between 10$^{12}$-10$^{16}$~s$^{-1}$, we obtain an adsorption
energy in the range 1.06-1.40~eV.  The difference between
experiment and vdW-DF energies is due to the uncertainty in the
GGA exchange energies \cite{Puzder2006a}, which can be seen in 
Table~\ref{tab:vdW-DF}.  The use of PBE with the vdW-DF increases the adsorption energies to
within the experimental range but the uncertainty in the exchange energy    
is large enough to mask the small energy differences between the two
structures, particularly for the lower coverages.  

In summary, we have demonstrated that standard DFT adsorption
energies are significantly dependent on the xc functional and,
furthermore, the inclusion of vdW forces makes a {\it qualitative}
difference to the results.  
Standard DFT with PBE and revPBE functionals finds that the TB
structure is always stable, whereas vdW-DF DFT calculations find
that, for some coverages, vdW forces stabilise the BF structure.  
These results have significant implications for many DFT studies
as vdW forces are generally considered to have a negligible effect on
covalently bonded systems and are usually ignored.  

KJ would like to thank Andris Gulans for valuable
discussions.  This work was supported by the
Finnish Funding Agency for Technology and Innovation (TEKES) and
the Academy of Finland (Center of Excellence 2006-2011).
Computational resources were provided by the
Center for Scientific Computing (CSC), Finland.  The Lundbeck
Foundation (Denmark) is gratefully acknowledged.

\bibliography{../../hypris}

\begin{thebibliography}{36}
\expandafter\ifx\csname natexlab\endcsname\relax\def\natexlab#1{#1}\fi
\expandafter\ifx\csname bibnamefont\endcsname\relax
  \def\bibnamefont#1{#1}\fi
\expandafter\ifx\csname bibfnamefont\endcsname\relax
  \def\bibfnamefont#1{#1}\fi
\expandafter\ifx\csname citenamefont\endcsname\relax
  \def\citenamefont#1{#1}\fi
\expandafter\ifx\csname url\endcsname\relax
  \def\url#1{\texttt{#1}}\fi
\expandafter\ifx\csname urlprefix\endcsname\relax\def\urlprefix{URL }\fi
\providecommand{\bibinfo}[2]{#2}
\providecommand{\eprint}[2][]{\url{#2}}

\bibitem[{\citenamefont{Hersam et~al.}(2000)\citenamefont{Hersam, Guisinger,
  and Lyding}}]{Hersam2000a}
\bibinfo{author}{\bibfnamefont{M.~C.} \bibnamefont{Hersam}},
  \bibinfo{author}{\bibfnamefont{N.~P.} \bibnamefont{Guisinger}},
  \bibnamefont{and} \bibinfo{author}{\bibfnamefont{J.~W.}
  \bibnamefont{Lyding}}, \bibinfo{journal}{Nanotechnology}
  \textbf{\bibinfo{volume}{11}}, \bibinfo{pages}{70} (\bibinfo{year}{2000}).

\bibitem[{\citenamefont{Bent}(2002)}]{Bent2002a}
\bibinfo{author}{\bibfnamefont{S.~F.} \bibnamefont{Bent}},
  \bibinfo{journal}{Surface Science} \textbf{\bibinfo{volume}{500}},
  \bibinfo{pages}{879} (\bibinfo{year}{2002}).

\bibitem[{\citenamefont{Duwez et~al.}(2006)\citenamefont{Duwez, Cuenot,
  J\'{e}r\^{o}me, Gabriel, J\'{e}r\^{o}me, Rapino, and Zerbetto}}]{Duwez2006a}
\bibinfo{author}{\bibfnamefont{A.-S.} \bibnamefont{Duwez}},
  \bibinfo{author}{\bibfnamefont{S.}~\bibnamefont{Cuenot}},
  \bibinfo{author}{\bibfnamefont{C.}~\bibnamefont{J\'{e}r\^{o}me}},
  \bibinfo{author}{\bibfnamefont{S.}~\bibnamefont{Gabriel}},
  \bibinfo{author}{\bibfnamefont{R.}~\bibnamefont{J\'{e}r\^{o}me}},
  \bibinfo{author}{\bibfnamefont{S.}~\bibnamefont{Rapino}}, \bibnamefont{and}
  \bibinfo{author}{\bibfnamefont{F.}~\bibnamefont{Zerbetto}},
  \bibinfo{journal}{Nature Nanotechnology} \textbf{\bibinfo{volume}{1}},
  \bibinfo{pages}{122} (\bibinfo{year}{2006}).

\bibitem[{\citenamefont{Zanella et~al.}(2006)\citenamefont{Zanella, Fazzio, and
  da~Silva}}]{Zanella2006a}
\bibinfo{author}{\bibfnamefont{I.}~\bibnamefont{Zanella}},
  \bibinfo{author}{\bibfnamefont{A.}~\bibnamefont{Fazzio}}, \bibnamefont{and}
  \bibinfo{author}{\bibfnamefont{A.~J.~R.} \bibnamefont{da~Silva}},
  \bibinfo{journal}{J. Phys. Chem. B} \textbf{\bibinfo{volume}{110}},
  \bibinfo{pages}{10849} (\bibinfo{year}{2006}).

\bibitem[{\citenamefont{Quek et~al.}(2007)\citenamefont{Quek, Neaton,
  Hybertsen, Kaxiras, and Louie}}]{Quek2007a}
\bibinfo{author}{\bibfnamefont{S.~Y.} \bibnamefont{Quek}},
  \bibinfo{author}{\bibfnamefont{J.~B.} \bibnamefont{Neaton}},
  \bibinfo{author}{\bibfnamefont{M.~S.} \bibnamefont{Hybertsen}},
  \bibinfo{author}{\bibfnamefont{E.}~\bibnamefont{Kaxiras}}, \bibnamefont{and}
  \bibinfo{author}{\bibfnamefont{S.~G.} \bibnamefont{Louie}},
  \bibinfo{journal}{Phys. Rev. Lett.} \textbf{\bibinfo{volume}{98}},
  \bibinfo{pages}{66807} (\bibinfo{year}{2007}).

\bibitem[{\citenamefont{\v{S}tich et~al.}(1992)\citenamefont{\v{S}tich, Payne,
  King-Smith, Lin, and Clarke}}]{Stich1992a}
\bibinfo{author}{\bibfnamefont{I.}~\bibnamefont{\v{S}tich}},
  \bibinfo{author}{\bibfnamefont{M.~C.} \bibnamefont{Payne}},
  \bibinfo{author}{\bibfnamefont{R.~D.} \bibnamefont{King-Smith}},
  \bibinfo{author}{\bibfnamefont{J.-S.} \bibnamefont{Lin}}, \bibnamefont{and}
  \bibinfo{author}{\bibfnamefont{L.~J.} \bibnamefont{Clarke}},
  \bibinfo{journal}{Phys. Rev. Lett.} \textbf{\bibinfo{volume}{68}},
  \bibinfo{pages}{1351} (\bibinfo{year}{1992}).

\bibitem[{\citenamefont{Cooper et~al.}(2007)\citenamefont{Cooper, Thonhauser,
  Puzder, Schr\"{o}der, Lundqvist, and Langreth}}]{Cooper2007b}
\bibinfo{author}{\bibfnamefont{V.~R.} \bibnamefont{Cooper}},
  \bibinfo{author}{\bibfnamefont{T.}~\bibnamefont{Thonhauser}},
  \bibinfo{author}{\bibfnamefont{A.}~\bibnamefont{Puzder}},
  \bibinfo{author}{\bibfnamefont{E.}~\bibnamefont{Schr\"{o}der}},
  \bibinfo{author}{\bibfnamefont{B.~I.} \bibnamefont{Lundqvist}},
  \bibnamefont{and} \bibinfo{author}{\bibfnamefont{D.~C.}
  \bibnamefont{Langreth}}, \bibinfo{journal}{J. Am. Chem. Soc.}
  \textbf{\bibinfo{volume}{130}}, \bibinfo{pages}{1304} (\bibinfo{year}{2007}).

\bibitem[{\citenamefont{Gokhale et~al.}(1998)\citenamefont{Gokhale,
  Trischberger, Menzel, Widdra, Dr\"{o}ge, Steinr\"{u}ck, Birkenheuer,
  Gutdeutsch, and R\"{o}sch}}]{Gokhale1998a}
\bibinfo{author}{\bibfnamefont{S.}~\bibnamefont{Gokhale}},
  \bibinfo{author}{\bibfnamefont{P.}~\bibnamefont{Trischberger}},
  \bibinfo{author}{\bibfnamefont{D.}~\bibnamefont{Menzel}},
  \bibinfo{author}{\bibfnamefont{W.}~\bibnamefont{Widdra}},
  \bibinfo{author}{\bibfnamefont{H.}~\bibnamefont{Dr\"{o}ge}},
  \bibinfo{author}{\bibfnamefont{H.~P.} \bibnamefont{Steinr\"{u}ck}},
  \bibinfo{author}{\bibfnamefont{U.}~\bibnamefont{Birkenheuer}},
  \bibinfo{author}{\bibfnamefont{U.}~\bibnamefont{Gutdeutsch}},
  \bibnamefont{and}
  \bibinfo{author}{\bibfnamefont{N.}~\bibnamefont{R\"{o}sch}},
  \bibinfo{journal}{J. Chem. Phys.} \textbf{\bibinfo{volume}{108}},
  \bibinfo{pages}{5554} (\bibinfo{year}{1998}).

\bibitem[{\citenamefont{Witkowski et~al.}(2003)\citenamefont{Witkowski,
  Hennies, Pietzsch, Mattsson, F\"{o}hlisch, Wurth, Nagasono, and
  Piancastelli}}]{Witkowski2003a}
\bibinfo{author}{\bibfnamefont{N.}~\bibnamefont{Witkowski}},
  \bibinfo{author}{\bibfnamefont{F.}~\bibnamefont{Hennies}},
  \bibinfo{author}{\bibfnamefont{A.}~\bibnamefont{Pietzsch}},
  \bibinfo{author}{\bibfnamefont{S.}~\bibnamefont{Mattsson}},
  \bibinfo{author}{\bibfnamefont{A.}~\bibnamefont{F\"{o}hlisch}},
  \bibinfo{author}{\bibfnamefont{W.}~\bibnamefont{Wurth}},
  \bibinfo{author}{\bibfnamefont{M.}~\bibnamefont{Nagasono}}, \bibnamefont{and}
  \bibinfo{author}{\bibfnamefont{M.~N.} \bibnamefont{Piancastelli}},
  \bibinfo{journal}{Phys. Rev. B} \textbf{\bibinfo{volume}{68}},
  \bibinfo{pages}{115408} (\bibinfo{year}{2003}).

\bibitem[{\citenamefont{Witkowski et~al.}(2005)\citenamefont{Witkowski,
  Pluchery, and Borensztein}}]{Witkowski2005a}
\bibinfo{author}{\bibfnamefont{N.}~\bibnamefont{Witkowski}},
  \bibinfo{author}{\bibfnamefont{O.}~\bibnamefont{Pluchery}}, \bibnamefont{and}
  \bibinfo{author}{\bibfnamefont{Y.}~\bibnamefont{Borensztein}},
  \bibinfo{journal}{Phys. Rev. B} \textbf{\bibinfo{volume}{72}},
  \bibinfo{pages}{75354} (\bibinfo{year}{2005}).

\bibitem[{\citenamefont{Lopinski
  et~al.}(1998{\natexlab{a}})\citenamefont{Lopinski, Moffatt, and
  Wolkow}}]{Lopinski1998a}
\bibinfo{author}{\bibfnamefont{G.~P.} \bibnamefont{Lopinski}},
  \bibinfo{author}{\bibfnamefont{D.~J.} \bibnamefont{Moffatt}},
  \bibnamefont{and} \bibinfo{author}{\bibfnamefont{R.~A.}
  \bibnamefont{Wolkow}}, \bibinfo{journal}{Chem. Phys. Lett.}
  \textbf{\bibinfo{volume}{282}}, \bibinfo{pages}{305}
  (\bibinfo{year}{1998}{\natexlab{a}}).

\bibitem[{\citenamefont{Lopinski
  et~al.}(1998{\natexlab{b}})\citenamefont{Lopinski, Fortier, Moffatt, and
  Wolkow}}]{Lopinski1998b}
\bibinfo{author}{\bibfnamefont{G.~P.} \bibnamefont{Lopinski}},
  \bibinfo{author}{\bibfnamefont{T.~M.} \bibnamefont{Fortier}},
  \bibinfo{author}{\bibfnamefont{D.~J.} \bibnamefont{Moffatt}},
  \bibnamefont{and} \bibinfo{author}{\bibfnamefont{R.~A.}
  \bibnamefont{Wolkow}}, \bibinfo{journal}{J. Vac. Sci. Technol. A}
  \textbf{\bibinfo{volume}{16}}, \bibinfo{pages}{1037}
  (\bibinfo{year}{1998}{\natexlab{b}}).

\bibitem[{\citenamefont{Kim et~al.}(2005)\citenamefont{Kim, Lee, and
  Yeom}}]{Kim2005a}
\bibinfo{author}{\bibfnamefont{Y.~K.} \bibnamefont{Kim}},
  \bibinfo{author}{\bibfnamefont{M.~H.} \bibnamefont{Lee}}, \bibnamefont{and}
  \bibinfo{author}{\bibfnamefont{H.~W.} \bibnamefont{Yeom}},
  \bibinfo{journal}{Phys. Rev. B} \textbf{\bibinfo{volume}{71}},
  \bibinfo{pages}{115311} (\bibinfo{year}{2005}).

\bibitem[{\citenamefont{Hofer et~al.}(2001)\citenamefont{Hofer, Fisher,
  Lopinski, and Wolkow}}]{Hofer2001a}
\bibinfo{author}{\bibfnamefont{W.~A.} \bibnamefont{Hofer}},
  \bibinfo{author}{\bibfnamefont{A.~J.} \bibnamefont{Fisher}},
  \bibinfo{author}{\bibfnamefont{G.~P.} \bibnamefont{Lopinski}},
  \bibnamefont{and} \bibinfo{author}{\bibfnamefont{R.~A.}
  \bibnamefont{Wolkow}}, \bibinfo{journal}{Phys. Rev. B}
  \textbf{\bibinfo{volume}{63}}, \bibinfo{pages}{85314} (\bibinfo{year}{2001}).

\bibitem[{\citenamefont{Lee and Cho}(2005)}]{Lee2005a}
\bibinfo{author}{\bibfnamefont{J.-Y.} \bibnamefont{Lee}} \bibnamefont{and}
  \bibinfo{author}{\bibfnamefont{J.-H.} \bibnamefont{Cho}},
  \bibinfo{journal}{Phys. Rev. B} \textbf{\bibinfo{volume}{72}},
  \bibinfo{pages}{235317} (\bibinfo{year}{2005}).

\bibitem[{\citenamefont{Mamatkulov et~al.}(2006)\citenamefont{Mamatkulov,
  Stauffer, Minot, and Sonnet}}]{Mamatkulov2006a}
\bibinfo{author}{\bibfnamefont{M.}~\bibnamefont{Mamatkulov}},
  \bibinfo{author}{\bibfnamefont{L.}~\bibnamefont{Stauffer}},
  \bibinfo{author}{\bibfnamefont{C.}~\bibnamefont{Minot}}, \bibnamefont{and}
  \bibinfo{author}{\bibfnamefont{P.}~\bibnamefont{Sonnet}},
  \bibinfo{journal}{Phys. Rev. B} \textbf{\bibinfo{volume}{73}},
  \bibinfo{pages}{35321} (\bibinfo{year}{2006}).

\bibitem[{\citenamefont{Jung and Gordon}(2005)}]{Jung2005a}
\bibinfo{author}{\bibfnamefont{Y.}~\bibnamefont{Jung}} \bibnamefont{and}
  \bibinfo{author}{\bibfnamefont{M.~S.} \bibnamefont{Gordon}},
  \bibinfo{journal}{J. Am. Chem. Soc.} \textbf{\bibinfo{volume}{127}},
  \bibinfo{pages}{3131} (\bibinfo{year}{2005}).

\bibitem[{\citenamefont{Dion et~al.}(2004)\citenamefont{Dion, Rydberg,
  Schr\"{o}der, Langreth, and Lundqvist}}]{Dion2004a}
\bibinfo{author}{\bibfnamefont{M.}~\bibnamefont{Dion}},
  \bibinfo{author}{\bibfnamefont{H.}~\bibnamefont{Rydberg}},
  \bibinfo{author}{\bibfnamefont{E.}~\bibnamefont{Schr\"{o}der}},
  \bibinfo{author}{\bibfnamefont{D.~C.} \bibnamefont{Langreth}},
  \bibnamefont{and} \bibinfo{author}{\bibfnamefont{B.~I.}
  \bibnamefont{Lundqvist}}, \bibinfo{journal}{Phys. Rev. Lett.}
  \textbf{\bibinfo{volume}{92}}, \bibinfo{pages}{246401}
  (\bibinfo{year}{2004}).

\bibitem[{\citenamefont{Puzder et~al.}(2006)\citenamefont{Puzder, Dion, and
  Langreth}}]{Puzder2006a}
\bibinfo{author}{\bibfnamefont{A.}~\bibnamefont{Puzder}},
  \bibinfo{author}{\bibfnamefont{M.}~\bibnamefont{Dion}}, \bibnamefont{and}
  \bibinfo{author}{\bibfnamefont{D.~C.} \bibnamefont{Langreth}},
  \bibinfo{journal}{J. Chem. Phys.} \textbf{\bibinfo{volume}{124}},
  \bibinfo{pages}{164105} (\bibinfo{year}{2006}).

\bibitem[{\citenamefont{Thonhauser et~al.}(2006)\citenamefont{Thonhauser,
  Puzder, and Langreth}}]{Thonhauser2006a}
\bibinfo{author}{\bibfnamefont{T.}~\bibnamefont{Thonhauser}},
  \bibinfo{author}{\bibfnamefont{A.}~\bibnamefont{Puzder}}, \bibnamefont{and}
  \bibinfo{author}{\bibfnamefont{D.~C.} \bibnamefont{Langreth}},
  \bibinfo{journal}{J. Chem. Phys.} \textbf{\bibinfo{volume}{124}},
  \bibinfo{pages}{164106} (\bibinfo{year}{2006}).

\bibitem[{\citenamefont{Thonhauser et~al.}(2007)\citenamefont{Thonhauser,
  Cooper, Li, Puzder, Hyldgaard, and Langreth}}]{Thonhauser2007a}
\bibinfo{author}{\bibfnamefont{T.}~\bibnamefont{Thonhauser}},
  \bibinfo{author}{\bibfnamefont{V.~R.} \bibnamefont{Cooper}},
  \bibinfo{author}{\bibfnamefont{S.}~\bibnamefont{Li}},
  \bibinfo{author}{\bibfnamefont{A.}~\bibnamefont{Puzder}},
  \bibinfo{author}{\bibfnamefont{P.}~\bibnamefont{Hyldgaard}},
  \bibnamefont{and} \bibinfo{author}{\bibfnamefont{D.~C.}
  \bibnamefont{Langreth}}, \bibinfo{journal}{Phys. Rev. B}
  \textbf{\bibinfo{volume}{76}}, \bibinfo{pages}{125112}
  (\bibinfo{year}{2007}).

\bibitem[{\citenamefont{Ziambaras et~al.}(2007)\citenamefont{Ziambaras, Kleis,
  Schr\"{o}der, and Hyldgaard}}]{Ziambaras2007a}
\bibinfo{author}{\bibfnamefont{E.}~\bibnamefont{Ziambaras}},
  \bibinfo{author}{\bibfnamefont{J.}~\bibnamefont{Kleis}},
  \bibinfo{author}{\bibfnamefont{E.}~\bibnamefont{Schr\"{o}der}},
  \bibnamefont{and}
  \bibinfo{author}{\bibfnamefont{P.}~\bibnamefont{Hyldgaard}},
  \bibinfo{journal}{Phys. Rev. B} \textbf{\bibinfo{volume}{76}},
  \bibinfo{pages}{155425} (\bibinfo{year}{2007}).

\bibitem[{\citenamefont{Chakarova-K\"{a}ck
  et~al.}(2006)\citenamefont{Chakarova-K\"{a}ck, Schr\"{o}der, Lundqvist, and
  Langreth}}]{Chakarova2006a}
\bibinfo{author}{\bibfnamefont{S.~D.} \bibnamefont{Chakarova-K\"{a}ck}},
  \bibinfo{author}{\bibfnamefont{E.}~\bibnamefont{Schr\"{o}der}},
  \bibinfo{author}{\bibfnamefont{B.~I.} \bibnamefont{Lundqvist}},
  \bibnamefont{and} \bibinfo{author}{\bibfnamefont{D.~C.}
  \bibnamefont{Langreth}}, \bibinfo{journal}{Phys. Rev. Lett.}
  \textbf{\bibinfo{volume}{96}}, \bibinfo{pages}{146107}
  (\bibinfo{year}{2006}).

\bibitem[{Dac()}]{Dacapo}
\emph{\bibinfo{title}{{Open source plane-wave DFT code DACAPO}}},
  \bibinfo{howpublished}{http://www.fysik.dtu.dk/CAMPOS/}.

\bibitem[{\citenamefont{Perdew et~al.}(1996{\natexlab{a}})\citenamefont{Perdew,
  Burke, and Ernzerhof}}]{Perdew1996a}
\bibinfo{author}{\bibfnamefont{J.~P.} \bibnamefont{Perdew}},
  \bibinfo{author}{\bibfnamefont{K.}~\bibnamefont{Burke}}, \bibnamefont{and}
  \bibinfo{author}{\bibfnamefont{M.}~\bibnamefont{Ernzerhof}},
  \bibinfo{journal}{Phys. Rev. Lett.} \textbf{\bibinfo{volume}{77}},
  \bibinfo{pages}{3865} (\bibinfo{year}{1996}{\natexlab{a}}).

\bibitem[{\citenamefont{Perdew et~al.}(1997)\citenamefont{Perdew, Burke, and
  Ernzerhof}}]{Perdew1997a}
\bibinfo{author}{\bibfnamefont{J.~P.} \bibnamefont{Perdew}},
  \bibinfo{author}{\bibfnamefont{K.}~\bibnamefont{Burke}}, \bibnamefont{and}
  \bibinfo{author}{\bibfnamefont{M.}~\bibnamefont{Ernzerhof}},
  \bibinfo{journal}{Phys. Rev. Lett.} \textbf{\bibinfo{volume}{78}},
  \bibinfo{pages}{1396(E)} (\bibinfo{year}{1997}).

\bibitem[{\citenamefont{Perdew et~al.}(1998)\citenamefont{Perdew, Burke, Zupan,
  and Ernzerhof}}]{Perdew1998a}
\bibinfo{author}{\bibfnamefont{J.~P.} \bibnamefont{Perdew}},
  \bibinfo{author}{\bibfnamefont{K.}~\bibnamefont{Burke}},
  \bibinfo{author}{\bibfnamefont{A.}~\bibnamefont{Zupan}}, \bibnamefont{and}
  \bibinfo{author}{\bibfnamefont{M.}~\bibnamefont{Ernzerhof}},
  \bibinfo{journal}{J. Chem. Phys.} \textbf{\bibinfo{volume}{108}},
  \bibinfo{pages}{1522} (\bibinfo{year}{1998}).

\bibitem[{\citenamefont{Zhang and Yang}(1998)}]{Zhang1998a}
\bibinfo{author}{\bibfnamefont{Y.}~\bibnamefont{Zhang}} \bibnamefont{and}
  \bibinfo{author}{\bibfnamefont{W.}~\bibnamefont{Yang}},
  \bibinfo{journal}{Phys. Rev. Lett.} \textbf{\bibinfo{volume}{80}},
  \bibinfo{pages}{890} (\bibinfo{year}{1998}).

\bibitem[{\citenamefont{Hammer et~al.}(1999)\citenamefont{Hammer, Hansen, and
  Norskov}}]{Hammer1999a}
\bibinfo{author}{\bibfnamefont{B.}~\bibnamefont{Hammer}},
  \bibinfo{author}{\bibfnamefont{L.~B.} \bibnamefont{Hansen}},
  \bibnamefont{and} \bibinfo{author}{\bibfnamefont{J.~K.}
  \bibnamefont{Norskov}}, \bibinfo{journal}{Phys. Rev. B}
  \textbf{\bibinfo{volume}{59}}, \bibinfo{pages}{7413} (\bibinfo{year}{1999}).

\bibitem[{\citenamefont{Perdew}(1991)}]{Perdew1991a}
\bibinfo{author}{\bibfnamefont{J.~P.} \bibnamefont{Perdew}},
  \emph{\bibinfo{title}{Electronic Structure of Solids '91}}
  (\bibinfo{publisher}{Academie Verlag}, \bibinfo{year}{1991}),
  p.~\bibinfo{pages}{11}.

\bibitem[{\citenamefont{Perdew et~al.}(1996{\natexlab{b}})\citenamefont{Perdew,
  Burke, and Wang}}]{Perdew1996b}
\bibinfo{author}{\bibfnamefont{J.~P.} \bibnamefont{Perdew}},
  \bibinfo{author}{\bibfnamefont{K.}~\bibnamefont{Burke}}, \bibnamefont{and}
  \bibinfo{author}{\bibfnamefont{Y.}~\bibnamefont{Wang}},
  \bibinfo{journal}{Phys. Rev. B} \textbf{\bibinfo{volume}{54}},
  \bibinfo{pages}{16533} (\bibinfo{year}{1996}{\natexlab{b}}).

\bibitem[{\citenamefont{Johnston and Nieminen}(2007)}]{Johnston2007a}
\bibinfo{author}{\bibfnamefont{K.}~\bibnamefont{Johnston}} \bibnamefont{and}
  \bibinfo{author}{\bibfnamefont{R.~M.} \bibnamefont{Nieminen}},
  \bibinfo{journal}{Phys. Rev. B} \textbf{\bibinfo{volume}{76}},
  \bibinfo{pages}{85402} (\bibinfo{year}{2007}).

\bibitem[{\citenamefont{Wu et~al.}(2001)\citenamefont{Wu, Vargas, Nayak,
  Lotrich, and Scoles}}]{Wu2001a}
\bibinfo{author}{\bibfnamefont{X.}~\bibnamefont{Wu}},
  \bibinfo{author}{\bibfnamefont{M.~C.} \bibnamefont{Vargas}},
  \bibinfo{author}{\bibfnamefont{S.}~\bibnamefont{Nayak}},
  \bibinfo{author}{\bibfnamefont{V.}~\bibnamefont{Lotrich}}, \bibnamefont{and}
  \bibinfo{author}{\bibfnamefont{G.}~\bibnamefont{Scoles}},
  \bibinfo{journal}{J. Chem. Phys.} \textbf{\bibinfo{volume}{115}},
  \bibinfo{pages}{8748} (\bibinfo{year}{2001}).

\bibitem[{\citenamefont{Langreth et~al.}(2005)\citenamefont{Langreth, Dion,
  Rydberg, Schr\"{o}der, Hyldgaard, and Lundqvist}}]{Langreth2005a}
\bibinfo{author}{\bibfnamefont{D.~C.} \bibnamefont{Langreth}},
  \bibinfo{author}{\bibfnamefont{M.}~\bibnamefont{Dion}},
  \bibinfo{author}{\bibfnamefont{H.}~\bibnamefont{Rydberg}},
  \bibinfo{author}{\bibfnamefont{E.}~\bibnamefont{Schr\"{o}der}},
  \bibinfo{author}{\bibfnamefont{P.}~\bibnamefont{Hyldgaard}},
  \bibnamefont{and} \bibinfo{author}{\bibfnamefont{B.~I.}
  \bibnamefont{Lundqvist}}, \bibinfo{journal}{Int. J. Quant. Chem.}
  \textbf{\bibinfo{volume}{101}}, \bibinfo{pages}{599} (\bibinfo{year}{2005}).

\bibitem[{\citenamefont{Kleis et~al.}(2007)\citenamefont{Kleis, Lundqvist,
  Langreth, and Schr\"{o}der}}]{Kleis2007a}
\bibinfo{author}{\bibfnamefont{J.}~\bibnamefont{Kleis}},
  \bibinfo{author}{\bibfnamefont{B.~I.} \bibnamefont{Lundqvist}},
  \bibinfo{author}{\bibfnamefont{D.~C.} \bibnamefont{Langreth}},
  \bibnamefont{and}
  \bibinfo{author}{\bibfnamefont{E.}~\bibnamefont{Schr\"{o}der}},
  \bibinfo{journal}{Phys. Rev. B} \textbf{\bibinfo{volume}{76}},
  \bibinfo{pages}{100201(R)} (\bibinfo{year}{2007}).

\bibitem[{\citenamefont{Redhead}(1962)}]{Redhead1962a}
\bibinfo{author}{\bibfnamefont{P.~A.} \bibnamefont{Redhead}},
  \bibinfo{journal}{Vacuum} \textbf{\bibinfo{volume}{12}}, \bibinfo{pages}{203}
  (\bibinfo{year}{1962}).

\end{thebibliography}

\end{document}